\def\X1859{XTE~J$1859$+$226$}

\def\rxte{{\it{RXTE}}}

\def\mnras{MNRAS}
\def\aap{A\&A}
\def\apj{ApJ}
\def\apjl{ApJ}
\def\apjs{ApJS}
\def\apss{Ap\&SS}
\def\araa{ARA\&A}
\def\iaucirc{IAU Circ.}

\documentclass{PoS}
\usepackage{amssymb}
\usepackage{natbib}
\usepackage{epsfig}
\usepackage{epsf}
\title{Are the two peaks of the Cathedral QPO real harmonics?}

\ShortTitle{Are the two peaks of the Cathedral QPO real harmonics?}

\author{\speaker{Jerome Rodriguez}\\
       CEA Service d'Astrophysique, Laboratoire AIM, Saclay, France\\
       E-mail: \email{jrodriguez@cea.fr}}

\author{Peggy Varni\`ere\\
        CNRS, Laboratoire APC, Paris, France\\}

\abstract{We present a  study of the two main peak of the so-called cathedral
QPO in XTE J1859+226. While looking at the temporal evolution of the
two features we show that they do not manifest the same amplitude of
variations of their power, and do not seem to follow the flux variations in the 
same way. We then present their RMS-spectra and show that they
do not have the same shape, slope and cut-off energy. 
We discuss these different facts and try to answer the question regarding the genuineness of
their harmonic relationship}

\FullConference{Fast X-ray timing and spectroscopy at extreme count rates\\
		 February 7-11, 2011\\
		 Champ\'ery, Switzerland}

\begin{document}

\section{Introduction}
One of the very actual question regarding the physics at work in black hole binaries  
\citep[BHB, a.k.a microquasars e.g.][for a review]{remillard06} concerns the orgin and nature 
of the quasi periodic oscillations (QPO) one can see while producing the power density 
spectra (PDS) of these sources. While low frequency QPOs (LFQPO)  have been commonly seen in almost 
all BHBs in their hardest states \citet{remillard06,homan05b}, high frequency QPOs have only been 
seen in a handful of them. In the study presented here we focus on LFQPOs. The latter have been further 
classified into  types A, B, or C based on their typical frequencies,  total RMS amplitude, time lags, and the 
overal shape of underlying continuum of the PDS  \citep[e.g.][]{remillard02, casella05}.  
We have recently proposed  a tentative classification of states based on the presence of the 
different types of QPOs \citep{varniere11}. \\
\indent Many models have been proposed to explain the origin and behaviour of these LFQPO, but none 
of them has thus far been able to explain all observational facts. It is, indeed, quite clear that the inner disk somehow 
 sets the frequency of LFQPOs  \citep[e.g.][]{muno99,rodrigue02_qpo,rodrigue02_aei, rodrigue04_1550, Mikles09}, but 
LFQPOs have high amplitudes in states dominated by emission at  hard X-rays, and recent studies have shown that 
their frequency is (also) correlated with the power law photon index \citep[e.g.][]{vignarca03, shaposhnikov07}. This could indicate  
a strong relation to the corona. Finally  the RMS-spectra of LFQPO is hard (it increases with the energy) but also presents  
a cut-off whose energy is  variable  \citep{rodrigue04_1915,rodrigue08_1915b}. \\
\indent A much less explored properties of these features is related to the presence, and behaviour of (sub-) harmonically related 
peaks in the PDS. These harmonic can, in some cases, have properties that differ significantly from those 
of the fundamental QPO. This is, for example, the case of the type B QPO. For the latter, the fundamental and harmonics show
opposite signs of their times lags, and also different shape of their RMS-spectra \citep{casella04, cui99,homan01,rao10}.
In this paper we study the particular case of the so-called "Cathedral" QPO seen in the microquasars XTE J1859+226 \citet{casella04}.
This letter is a summary of our recently accepted paper \citep{rodrigue11}. It is organised as follows: in the next section 
we give a brief introduction on XTE J1859+226 and its temporal 
properties. We first present the basic properties of the peaks (Sec. 3), and  then present their temporal evolution  (Sec. 4) and spectra (Sec. 5). 
In regards of these analysis we discuss the potential association of the two peaks in the last section.

\section{XTE J1859+226}
\begin{figure}[htbp]
\centering
\epsfig{file=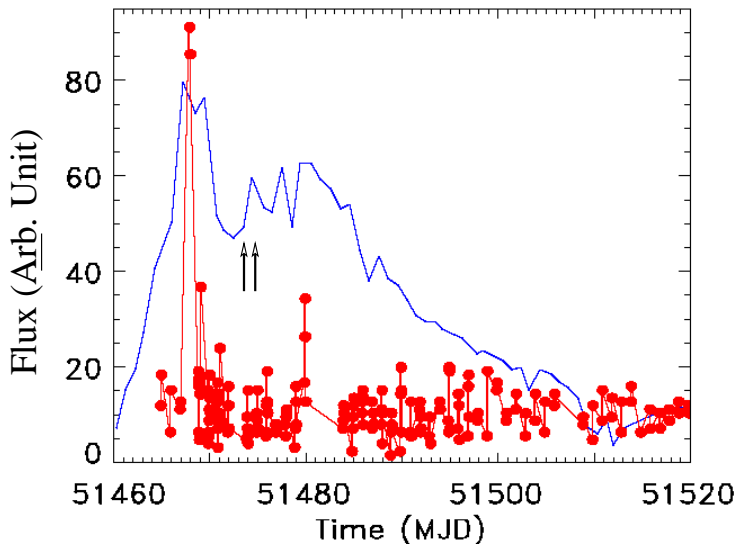, width=10cm}
\caption{RXTE/ASM (1.2-12 keV, blue) and Green Bank Interferometer (2.25 GHz, red) light curves of the 1999 outburst of XTE J1859+226. 
The two vertical arrows mark the position of the RXTE observations analysed in this paper. }
\label{fig:lc}
\end{figure}

\indent \X1859 was discovered on 1999 October 9 with the \rxte\ All Sky Monitor \citep{wood99} as it was entering into outburst. 
It is a microquasar given the observations of relativistic ejections in radio \citep{brocksopp02}. The RXTE/ASM and GBI 2.25 GHz 
light curves of the source are represented in Fig. \ref{fig:lc}.
\citet{cui00} observed LF and HFQPOs which led them to classify \X1859\ as a candidate BHB. An extensive 
timing analysis of this source is presented by \citet{casella04}.\X1859\ displays all three types of  LFQPOs, and, 
in two particular observations (on MJDs 51474.43 and 51475.43, Fig. \ref{fig:lc}), \citet{casella04} observed the 
presence of two peaks with harmonically related frequencies,  but unlike 
any other cases, similar RMS-amplitudes, that they dubbed  'Cathedral'-QPO.  Interestingly these authors remarked that the 
strongest peak (and highest in frequency, hereafter Peak 2)  has hard lags (the hard X-ray lag behind the soft X-rays), while  the lowest frequency 
peak (Peak 1) has soft lags. In this respect Peak 2 was considered as the fundamental, while Peak 1 was  the sub-harmonic. 

\section{Broad band fitting of the PDSs}
\begin{figure}[htbp]
\centering
\epsfig{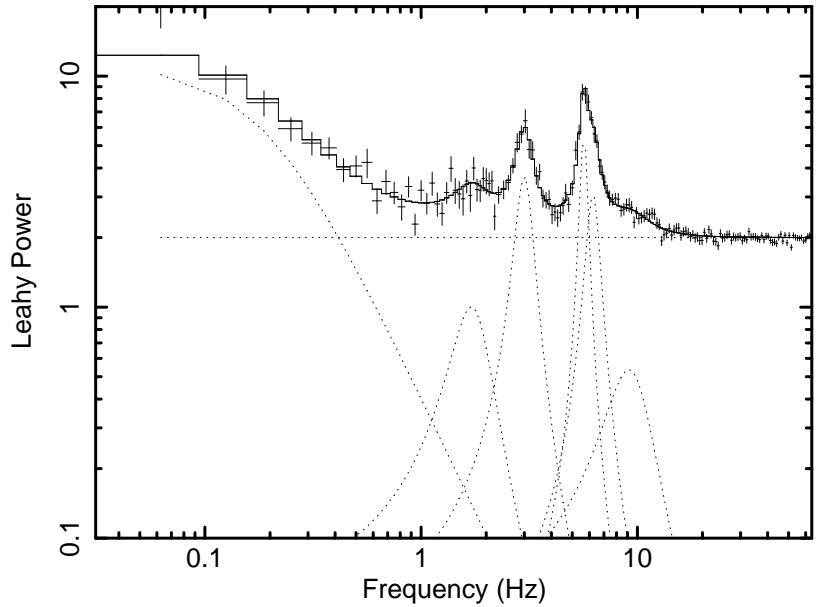}
\caption{Broad band fit of the PDS of the second observation. The dashed lines represent the different individual components.  }
\label{fig:QPO}
\end{figure}
The broad band PDSs of both observations were fitted with the additions of several Lorentzians \citep[see][ for the details 
of the data reduction and procedures of fittings]{rodrigue11}. We focus in the remaining of the two thin features at $\sim 3$ and $\sim 6$~Hz. Fig.~\ref{fig:QPO} shows 
the example of  the fit to the PDS of the second observation  (MJD 51475.43). In both observation Peak 2 may have a complex structure, and in fact it 
seems to be better represented by 2 thin Lorentzians (Fig.~\ref{fig:QPO} ). Note that \citet{casella04} also make a similar remark, but only in the 
case of the second observation. This additional feature is, however, poorly defined, and its parameters 
are badly constrained. We verify, by re-doing the whole analysis  that it had no significant impact on the 
other peaks, and since no influence was found it was omitted from our study, and is not further discussed here. \\
\indent The best peak parameters are the following for Obs. 1 (resp. Obs. 2) $\nu_1=2.94$~Hz, $Q_1=5.9$, $A_1=2.8\%$ (resp. $\nu_1=3.00$~Hz, 
$Q_1=5.2$, $A_1=2.9\%$), and  $\nu_2=5.83$ Hz, $Q_2=7.3$, $A_2=4.7\%$ (resp. $\nu_2=5.86$ Hz, $Q_2=6.5$, $A_2=4.6\%$).

\section{Temporal evolution}
\begin{figure}[htbp]
\centering
\epsfig{file=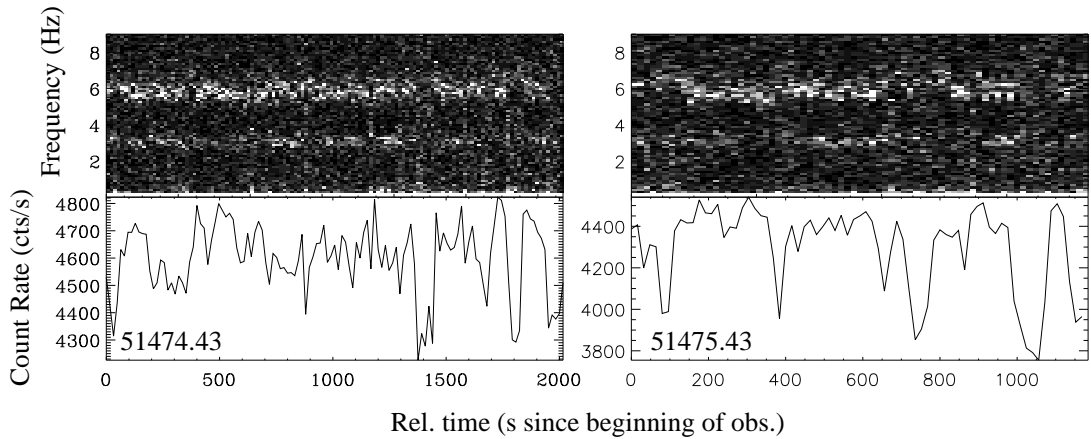, width=\columnwidth}
\caption{Dynamical PDS (top panels) and RXTE/PCA light curves of \X1859 during the two observations. Left: Obs. 1. Right: Obs.2  }
\label{fig:dynpo}
\end{figure}

We produced the dynamical PDS of \X1859\ to look to any  temporal evolution of the main peaks during each 
of the observations (Fig. ~\ref{fig:dynpo}). Both observation show similar temporal dependences of the two peaks. 
In both cases Peak 1 seems, on average, much weaker than Peak 2. It is strong only when the count rate is around 
its mean value. It is, in particular quite weak during the small flares, and is not visible during the dips. 
Peak 2, on the other hand, seems, in term of  power, more stable and seems to vary significantly only during the dips, where it 
may disappear. 
In fact a proper study of the dependences of the properties of the two peaks with the count rate \citep{rodrigue11} indeed shows that
both features undergo very different evolution with the count rate. The amplitude of Peak 1 decreases significantly with increasing count 
rate, while, at the same time, the amplitude of Peak 2 may show a linear increase.

\section{Energy dependences}
\begin{figure}[htbp]
\centering
\epsfig{file=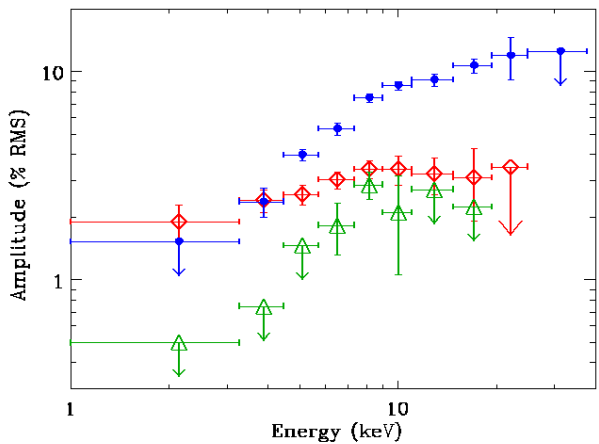, width=8cm}
\caption{RMS spectra obtained during Obs. 1 for Peak 1 (red) Peak 2 (blue)  and for the additional Lorentzian used to better represent Peak 2 (green). }
\label{fig:qpospec}
\end{figure}

The energy dependences of the peak's RMSs (in other words the RMS-Spectra) are reported in Fig.~\ref{fig:qpospec} for 
Peaks 1, 2, and the additional thin Lorentzian added to better represent Peak 2 (Sec. 3). Note, that, as mentioned in Sec. 3, 
the shapes of the RMS-Spectra of Peak 1 and 2 are the same if the third Lorentzian is not considered in the fits \citep{rodrigue11}.

It is pretty clear from Fig.\ref{fig:qpospec} that the two main peaks show a quite different behaviour. Peak 2 has a steeper (harder) 
spectrum than Peak 1. The latter first increases up to $\sim5.7$ keV and is then flat
until $\sim 20$~keV. Peak 2  is undetectable in the first energy bin (and is thus fainter than Peak 1). It 
increases up to $\sim20$~keV where its plateau is reached. 

\section{Conclusions}
While peaks 1 and 2 have frequencies that are in a harmonical relationship, they do not share the same properties.
They have opposite signs of their time lags \citep{casella04}, their temporal evolution is clearly different (Sec. 3), and their spectra have different slopes and cut-off energies (Sec. 5). 
The first tempting conclusion one could draw out of these results is that the integer factor between the two 
frequencies is fortuitous and the peaks are not related. This interpretation is, however, difficult to reconcile with the fact 
that the same type of QPO is commonly seen with harmonics. In addition the two observations presented here 
are separated by an observation showing another type of QPO, which makes it difficult to believe that in both 
the same fortuitous phenomenon occurred. \\
\indent The difference of spectral shape, in particular the fact that Peak 2 is harder, may indicate that the signal 
giving birth to the QPO is more sinusoidal at high energies \citep[][in the case of XTE J1550$-$564]{rao10}. This, however, does not account for the 
different signs of the time lags. In addition \citet{rao10} also mention that in fact the lowest frequency peak could be the fundamental. 
In that case the 'more sinusoidal' interpretation does not hold. \\
\indent All this suggest that, although some common physics might set the frequencies of the peaks, the origin of the two QPO is distinct.
Their different dependence on the count rate may indicate a kind of competing mechanism. This could be the case, for example, if the two 
peaks represent different modes of the same physical mechanism, that would be favoured at different moments.

\end{document}